# AI Augmented Digital Metal Component

Eunhyeok Seo[a*], Hyokyung Sung[b*], Hayeol Kim[a], Taekyeong Kim[a], Sangeun Park[b], Min Sik Lee[a], Seung Ki Moon[c], Jung Gi Kim[b], Hayoung Chung[a], Seong-Kyum Choi[d], Ji-hun Yu[e], Kyung Tae Kim[e], Seong Jin Park[f], Namhun Kim[a], Im Doo Jung[a†]

[a]Department of Mechanical Engineering,
Ulsan National Institute of Science and Technology, Ulsan 44919, Republic of Korea
[b]Department of Materials Engineering and Convergence Technology,
Gyeongsang National University, Jinju 52828, Republic of Korea
[c]School of Mechanical and Aerospace Engineering,
Nanyang Technological University, 50 Nanyang Ave, 639798, Singapore
[d]G.W.W. School of Mechanical Engineering, Georgia Institute of Technology, Atlanta, GA 30332, USA
[e]Powder & Ceramic Materials Division, Korea Institute of Materials Science, Changwon 51508, Republic of Korea
[f]Department of Mechanical Engineering, Pohang University of Science and Technology, Pohang 37673, Republic of Korea

## ABSTRACT

The aim of this work is to propose a new paradigm that imparts intelligence to metal parts with the fusion of metal additive manufacturing and artificial intelligence (AI). Our digital metal part classifies the status with real time data processing with convolutional neural network (CNN). The training data for the CNN is collected from a strain gauge embedded in metal parts by laser powder bed fusion process. We implement this approach using additive manufacturing, demonstrate a self-cognitive metal part recognizing partial screw loosening, malfunctioning, and external impacting object. The results indicate that metal part can recognize subtle change of multiple fixation state under repetitive compression with 89.1% accuracy with test sets. The proposed strategy showed promising potential in contributing to the hyper-connectivity for next generation of digital metal based mechanical systems

**Keywords:** 3D Printing, Artificial Intelligence, Laser Powder Bed Fusion, Convolutional Neural Network

*Equally contributed first author
†Corresponding author
E-Mail: idjung@unist.ac.kr (I.D. Jung, UNIST)

**I. INTRODUCTION**

The Fourth Industrial Revolution aims to overcome existing industrial limitations and bring about innovative changes through hyper-connectivity and super-intelligent technology. In the manufacturing industry, sensors, semiconductors, artificial intelligence, and Internet of Things (IoT) are combined to develop technologies that would have otherwise been impossible to establish using only existing manufacturing processes [1–5]. AI and IoT technologies play an important role in the development of intelligent manufacturing technologies such as smart factories, self-driving cars, and unmanned aerial vehicles through the real-time remote monitoring of mechanical systems [6–11].

Advancements in various artificial neural network techniques, including deep neural networks (DNNs) and convolutional neural networks (CNNs), have enabled AI technology to be used for several activities that cannot be accomplished by humans. In particular, AI technology can be used to determine and predict the state of mechanical systems [12,13]. Artificial neural networks have made considerable developments in media based on image and voice big data; they have also recently been expanding to various fields such as medicine and education. However, studies incorporating AI in the manufacturing field are lacking because it is difficult to collect meaningful big data in the physical environments of mechanical systems or factories [14]. Nevertheless, research is being conducted to apply AI technology by combining sensors and semiconductors with existing mechanical systems and factories [15,16]. Achieving connectivity and intelligence by combining sensors and semiconductors in manufacturing system has remained an important challenge.

A recent study aimed to combine a mechanical system and a sensor by attaching the sensor to the surface of the mechanical system; the collected data were monitored or used with AI technology to determine and predict the state of the system [17,18]. For example, a refinery for extracting oils, such as gasoline or diesel, uses an oil-refinement system built across a large area. Owing to the deterioration of the pumps of the oil-refinement system, the overall oil-refining efficiency decreased, which was a major issue. To this end, data collected through a vibration sensor attached to the pump surface were used with AI technology to predict when the pump components should be replaced [19,20]. However, analyses using physical data from the surface of a mechanical system have certain limitations when considering inner systems and sensors on the surface are easily degraded by the environment where the metal components are exposed. The attached sensor can negatively affect the performance of the mechanical system. For example, in the case of pressure tubes, the surface sensor does not indicate the internal pressure created by the internal fluid, also in the case of wind turbines, changes in wind flow due to surface sensors can alter aerodynamic performance [21,22]. While surface sensors can be lost by environmental factors, embedded sensors enable real-time monitoring at critical locations and can be isolated and protected from external environmental influences by the structure itself [23,24].

Additive manufacturing technology has been extensively developed in recent years; it is used to manufacture components of various materials and shapes [25–27]. Studies have recently been conducted to develop a technology that can diagnose the state of a component through semiconductors and sensors embedded into the component using additive manufacturing [28,29]. Such technologies that can embed sensors and semiconductors to additively manufacture polymer materials with low manufacturing temperatures have been actively developed, and recent studies have extended them to include materials such as metals and ceramics [30–32]. A technology that can embed sensors into existing metal-based systems and collect data to monitor and diagnose the state of each component is expected to contribute to hyper-connectivity and super-intelligent manufacturing systems.

In our study, a semiconductor is embedded inside a t-shaped metal component using metal additive manufacturing technology, and a CNN system was developed to determine the state of the component using sensor data collected in real time. A strain gauge was embedded inside the component through laser powder bed fusion (L-PBF), and the data collected in real time were applied to the CNN after performing fast Fourier transform (FFT) and image processing. The CNN was used to diagnose and predict the state and malfunctioning of the component, and the results were expressed by t-stochastic neighbor embedding (t-SNE).

## II. MATERIAL & METHODS

### 2.1 Sensor Selection & Sample Fabrication

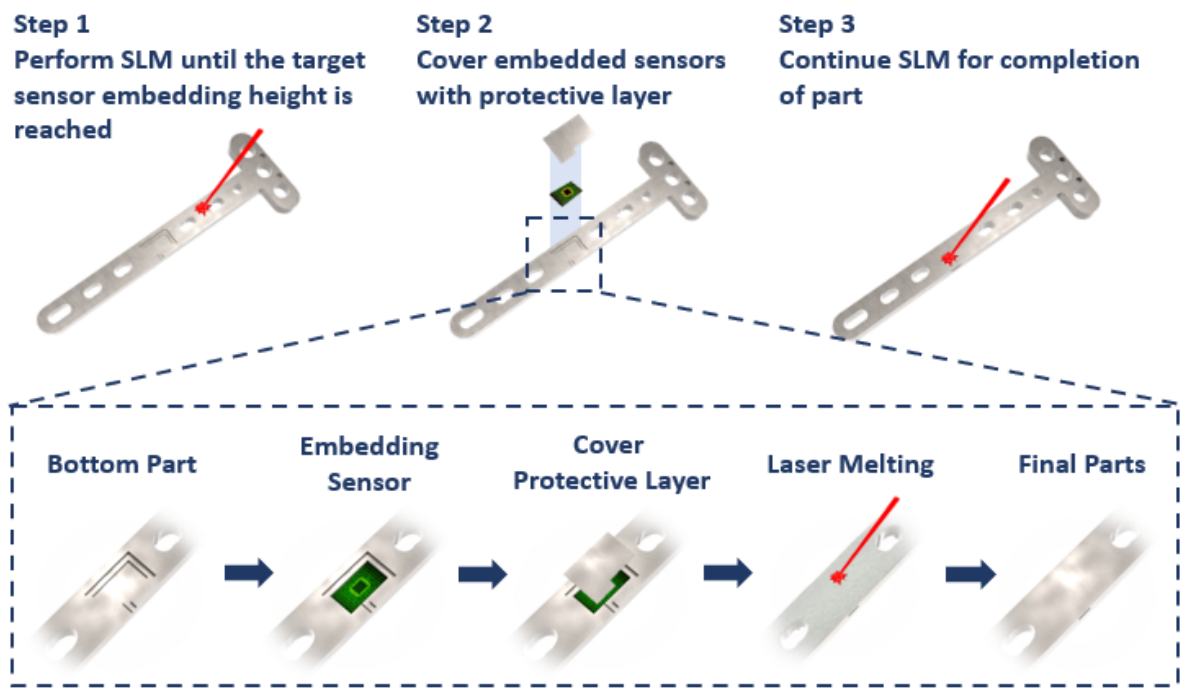

**Fig. 1. Schematic of L-PBF process for the AI augmented metal component.** (B) This L-PBF method consists of 3 steps. The bottom part is printed with the sensor embedding area. The sensor is located in the area, and after covering it with the protective layer, the manufacturing is completed by continuing the L-PBF process

For the compatibility with a circuit board (Arduino) that convert sensor data, strain gauges (BF350-3AA) were used with analog output (0~3.5V), nominal resistance of 350Ω, tolerance of <±0.1%, strain limit of 2%. M-Lab (Concept Laser, 2016) was used with a laser power of 90 W, scan speed of 800 mm/s, hatch space of 80 mm, layer thickness of 25 μm, spot size of 50 μm and an overlap of 30% for SUS316L (D50 size: 29.39 μm) powders in the L-PBF process. To ensure excellent mechanical properties, optimized manufacturing parameters from previous studies were used [33,34].

The metal components with dimensions of 105mm x 36mm x 5mm were prepared for the repeated compression test. The detailed dimensions of T-shaped metal components are shown in Fig. S1. For the embedding of the sensor during metal additive manufacturing, it was important to prepare a thermal protective layer which could prevent thermal degradation of sensor by laser processing [33]. We designed minimal space for strain gauge in the optimized location with thermal protective cover assembly place. The metal component consists of a first L-PBF part with this designed area where a sensor or IC chip can be inserted, and a second L-PBF part for finalizing the rest of the part. The protective layer was manufactured in a size suitable for the embedding area with same material of first & second L-PBF parts. Our L-PBF process consists of three main steps (**Fig. 1**). STEP 1: Print the first L-PBF part up to the height of the position where the sensor will be embedded. In this step, the L-PBF part including the sensor embedding area is manufactured. In our work, since the strain gauge is wired, we designed the L-PBF part to have a hole on the side through which the gauge's wire could exit. STEP 2: Remove the powder from the embedding area and place a strain gauge there. Cover the sensor with a pre-prepared protective layer and flatten the top surface for the second L-PBF part. In this process, a recoater is used to cover the new powder layer. STEP 3: Fabricate the remaining part based on the design of the second L-PBF part. The powder layer accumulated by the recoater is melted with laser to continue the L-PBF process. The metal component was completed by placing the sensor in the embedding area, covering it with the protective layer, and melting the new powder layer. Thus, using this special L-PBF process, we fabricated a metal component with embedded sensors that can generate physical deformation data.

*2.2 Topology Optimization*

For the topology optimization process, commercial software by Altair® Inspire was used to perform this topology optimization while SolidWorks® was used for finite element load prediction. In simulation, loading conditions were applied from the top and bottom of the T-shaped intelligent metal part. As a result of optimization of 65%, the removable part that does not significantly affect the mechanical properties was identified, and the part was determined as the optimal part for sensor embedding.

*2.3 Experimental Methods*

*2.3.1 Electron microscopy*

To investigate the microstructure of the metal component, electron backscatter diffraction (EBSD) and electron channeling contrast imaging (ECCI) analyses were conducted using a field-emission scanning electron microscope (JEOL, JSM-7610F). The EBSD was operated at 20 kV acceleration voltage, 20 mm working distance, and 0.25 μm step size. Kikuchi diffraction patterns were imaged by a EBSD detector (Oxford Instruments Symmetry®) and processed by the Oxford Instruments AZtec 2.0 EBSD software. To maximize the backscattered electron intensity, a tilted 70-degree angle was set between the specimen surface and the normal incidence of the electron beam. The acquired data was analyzed with the TSL OIM™ software.

ECC images were obtained by backscattered electron (BSE) Detector. ECCI observations were performed at 30 kV acceleration voltage, 2 nA probe current, and 15 mm working distance.

*2.3.2 Static Mechanical Testing*

DIC tensile test performed using plate-type tensile specimens (gauge length 12.5 mm, gauge width 3.2 mm, gauge thickness 0.7 mm). As there was no gauge length, two holes were tightened to conduct the tensile test. The DIC tensile specimen was connected to the grip in the same way as in the repeated compression experiment. The uniaxial tensile tests were conducted using a universal testing machine (MINOS-100, MTDI, Korea) with a 5M CCD camera, polarizer filter, blue light. The load cell (UT - 100F, MTDI, Korea) was used to measure the load. The test was performed at room temperature with a quasi-static strain rate of $1 \times 10^{-3}$ s$^{-1}$. The strain feedback compensation method was also used for the strain feedback control. The local strain evolution during the tensile test was measured using digital image correlation (DIC: ARAMIS 5M, GOM Optics, Germany) with black and white speckle patterns on the surface of the tensile specimens [31]. A single noncontact measurement system is utilized in DIC to measure the gage strain of the specimen. The DIC image was obtained at a frame rate of 1 Hz.

*2.3.3 Dynamic Mechanical testing*

To obtain deformation data of the metal component, a repeated compression tester (5944 Universal Testing Systems, Instron Corp., MA, USA) was used for 300 cycles with an initial load of 10 N and a displacement size of 0.35 mm. Two silicone columns, 80 mm in height and 52 mm in diameter, were connected vertically to the T-bone plate through screws. The test was conducted under four different settings: one normal state and three abnormal states. In the normal state, all screws are fully fastened; in abnormal state 1, the left screw is loose; in abnormal state 2, the left screw is missing; and in abnormal state 3, the left screw is loose, and the right screw is missing.

*2.4 Data featurization for machine learning*

The input factors in the state classification test are amplitude and time data which were obtained from strain gauge. The output factors consist of four types of statement: normal, abnormal 1, abnormal 2, and abnormal 3. For the impact recognition test, input factors are also amplitude and time data, and the output factors consist of three types as hand, hammer, and spanner. The generated data were stored as Excel file for each type of tests with specific period of time, and the stored data are read in the machine learning algorithm for the data pre-process: FFT and stacking data into spectrograms. As parameters for FFT, sample point N was set to 20, sample spacing T was set to 1/10, and time length was set to 6 seconds. Raw data were converted into spectrograms with a size of 10x5 through a half-cut process after FFT. Each spectrogram was labeled (normal, abnormal and etc) and divided into 85% of training sets and 15% of test sets. Once the CNN learning process is complete, the real time test is conducted with storing sensor data in an Excel file, and the trained machine learning algorithm continuously reads the real-time updated Excel file to distinguish if they or normal state, abnormal state and impact kinds.

*2.5 CNN process modeling*

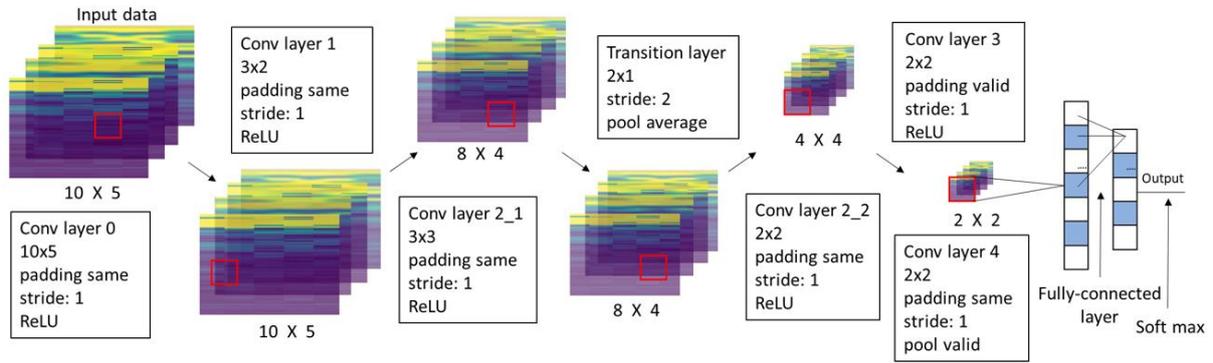

**Fig. 2. DenseNet-based convolutional neural network architecture.** DenseNet-based CNN was used to train the spectrogram and classify the states. Our CNN consists of convolutional layers and a transition layer that compute multiple feature maps. In the classification part, the fully connected layer and soft max flatten the image of 2x2 size and then normalize to a value between 0 and 1.

The optimized CNN model was based on DenseNet with six convolutional layers, one transition layer, a fully connected layer, and a softmax layer (**Fig. 2**). DenseNet prevents information from flowing between layers by using output information received from all the previous layers as input for the next layers. It also connects previous layers to the next layers to enhance feature propagation and reduce the number of parameters [33–35]. Between each convolutional layer, rectified linear unit (ReLU) functions were added as an activation function to reduce the vanishing gradient. A transition layer consisting of batch normalization, 1×1 convolution, and 2×2 average pooling was placed in the middle of the CNN model to reduce the width, vertical size, and number of feature maps. A 10×5 image was finally converted to a 2×2 image through the convolutional layers and the transition layer, and the states of the metal component were classified through the fully connected and softmax layers. The converted images were flattened with one layer via the fully connected layer, and then normalized to a value between 0 and 1 with the softmax layer. The training cost and the validation cost were compared according to learning rate in 8 cases (Fig. S8).

## III. RESULTS & DISCUSSION

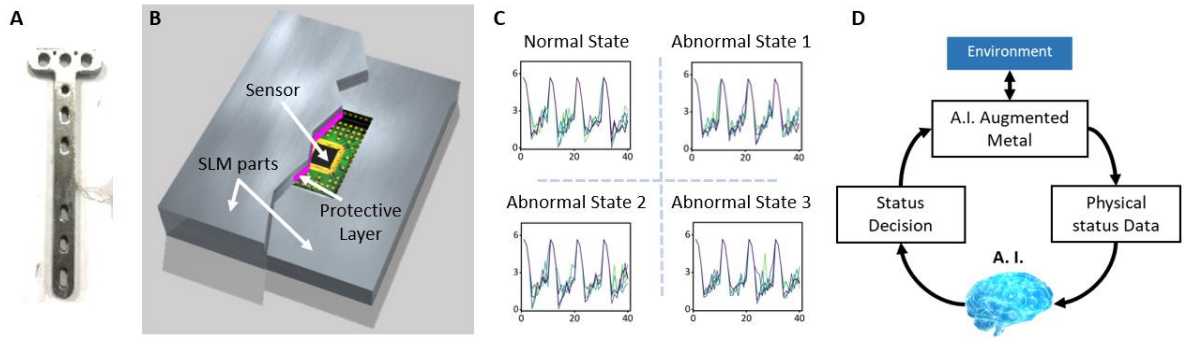

**Fig. 3. Overview of the AI augmented digital metal component.** (A, B) A strain gauge is embedded in the T-shaped AI-augmented metal with L-PBF process. (C) Four states of the metal component were assumed and used for classification. (D) This metal component can generate physical data due to the sensor, and the CNN acts like its brain to diagnose the condition.

Artificial intelligence augmented digital metal (AIADM) is the next generation of hyperconnected metal that will enable seamless monitoring of metal based mechanical system. To develop AIADM component, we integrated the physical data collecting system via additive manufacturing with a CNN classification. This intelligent metal is defined as a metal that prints physical data and classifies its state. The metal component was manufactured by the L-PBF method, and a strain gauge covered by a protective layer is embedded inside it (**Fig. 3, A and B**). As shown in **Fig. 3C**, the AIADM reads the change of environment and recognize its status as it collects physical big data and process the data with AI. The metal component status has four possible settings: one normal state and three abnormal states. In the normal state, all screws are fully fastened; in abnormal state 1, the left screw is loose; in abnormal state 2, the left screw is missing; and in abnormal state 3, the left screw is loose, and the right screw is also missing (**Fig. 3D**).

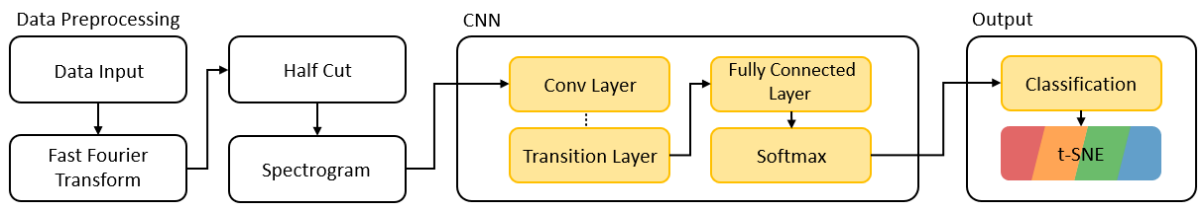

**Fig. 4. Metal component status classification flowchart.** The state of the metal component is classified as t-SNE after three main processes: data preprocessing, CNN, and t-SNE.

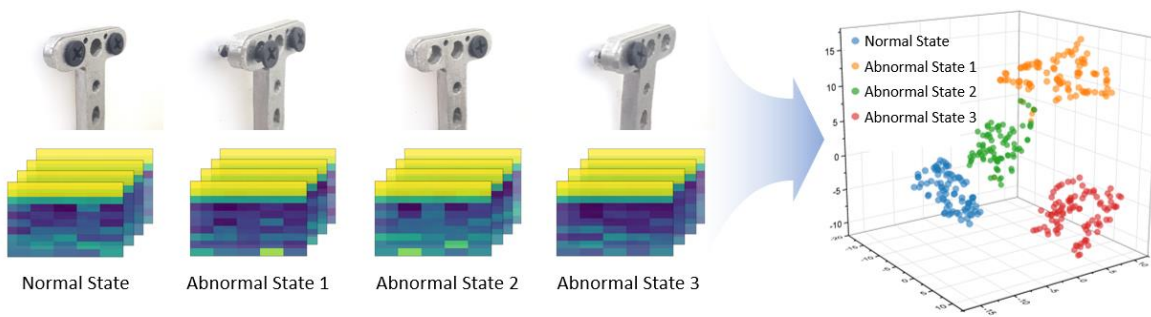

**Fig. 5. Visualization of classification results using t-SNE.** After learning spectrograms labeled with four states, the CNN classifies the results via t-SNE 3D plot.

The categorization of metal status process flow is shown in **Fig. 4**. First, in the data preprocessing part, the output strain data is fast Fourier transformed. The data are cut in half and converted into spectrograms. Spectrograms labeled with metal component status are used as the CNN's learning data. After being trained through the process of convolutional layer, transition layer, ReLU and batch normalization, fully connected layer in CNN, it is finally classified as a 3D visualized t-SNE graph. In **Fig. 5**, the data generated in each of the four states is converted into spectrograms which are two-dimensional compact representation of time series signal and the metal status are digitally visualized with 3D t-SNE after CNN process.

*3.1 Part integrity due to sensor embedding*

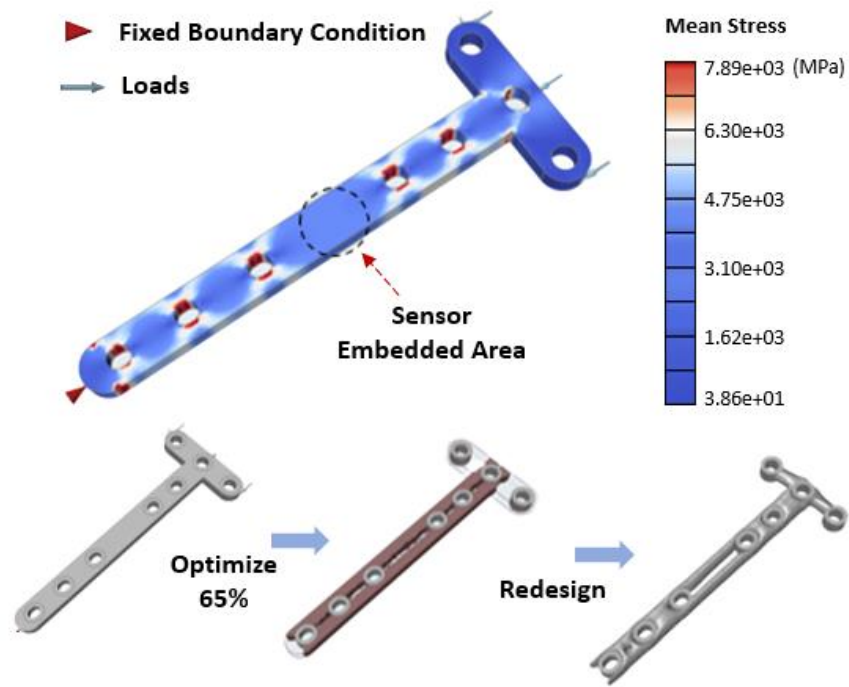

**Fig. 6. Device topology optimization.** As a result of deformation analysis after setting the boundary condition for the main load on the designed metal part, a lot of stress was applied around the hole. It was confirmed that even if a space was placed in the center by topology optimization, it did not significantly affect the decrease of the overall strength. Therefore, the sensor embedding area was determined as the center, which did not significantly affect the strength reduction.

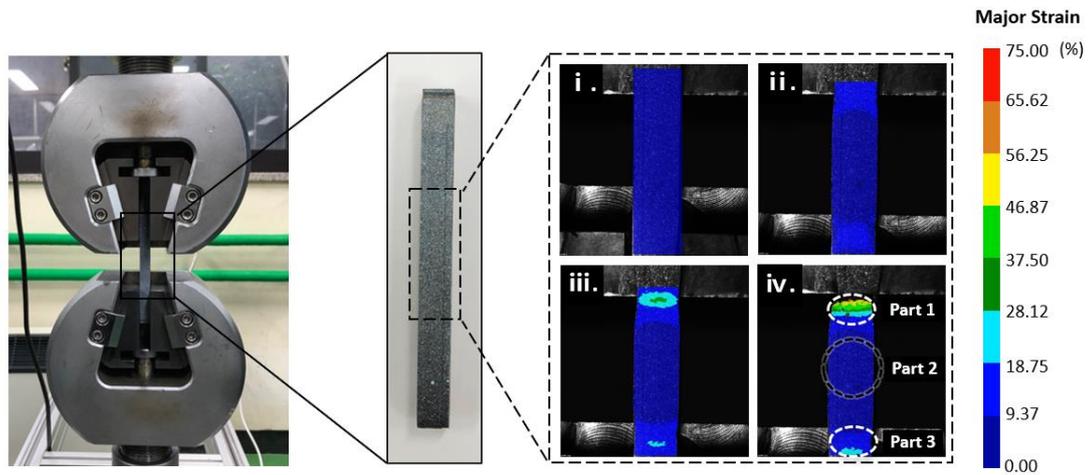

**Fig. 7. Results of the digital image correlation (DIC) tensile test.** The 15mm part above the sensor-embedded area was defined as Part 1, the sensor-embedded area was defined as Part 2, and the deformed area 17mm lower from the sensor-embedded area was defined as Part 3. The color change in Part 1 and Part 3 is distinct.

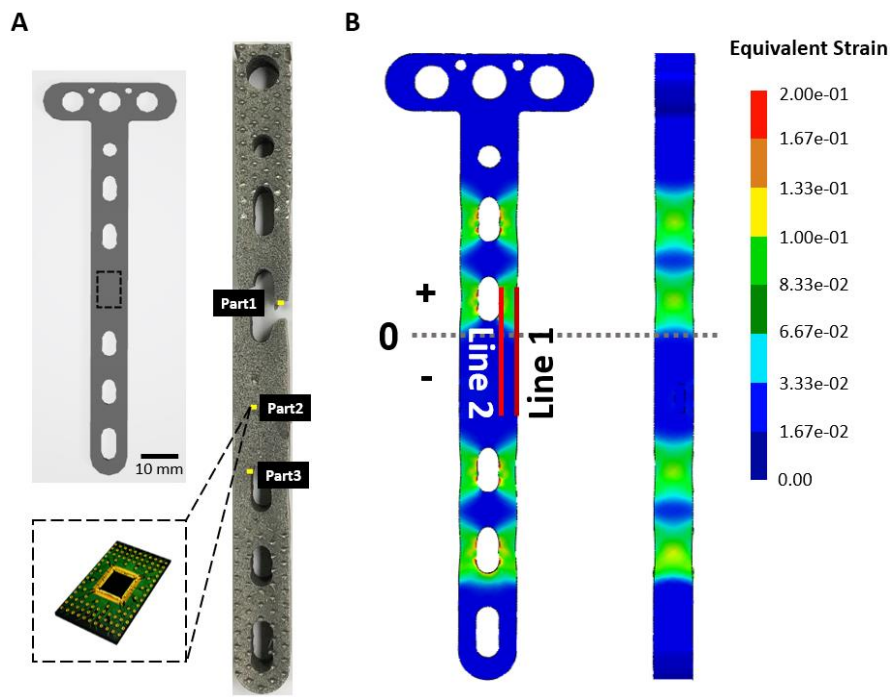

**Fig. 8. Tensile test and FEM analysis conducted to confirm the degradation of metal component properties due to sensor embedding.** (A) A tensile test was conducted by preparing a metal component tensile specimen. (B) FEM analysis was conducted on a sensor embedded metal component.

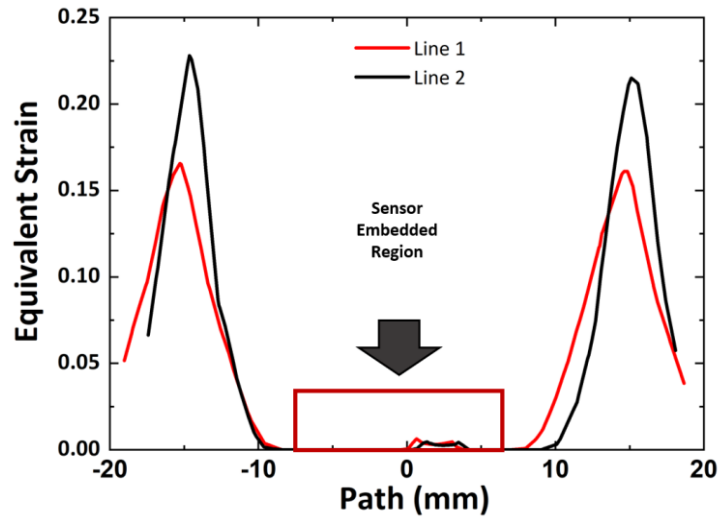

**Fig. 9. Equivalent strain result of FEM analysis.** In the FEM strain plot, the equivalent strain of the sensor embedded area and the periphery area was compared.

Unlike common metal components, our intelligent metal will be subjected to mechanical property degradation due to the sensor embedding process. In order to prevent this problem, the sensor embedding location was given priority consideration. To minimize the mechanical degradation of metal property, the optimum space for sensor embedding is decided with topology optimization (**Fig. 6**). After the topology optimization simulation, the designated center part could have vacant space with minimum mechanical property degradation, and it can be considered as the optimal position for embedding the sensor. A DIC tensile test (**Fig. 7**) and a finite element method (FEM) analysis (**Fig. 8**) were also performed to verify the mechanical property of sensor embedded t-shape metal parts. As shown in **Fig. 8A**, we prepared a tensile specimen of AIADM component and applied tension to it until fracture occurred. As the strain was enough to deform the metal part, the hole in the upper area (Part 1) was fractured. The center area (Part 2) where the sensor is embedded did not deform until the upper part was fractured. The lower part (Part 3), where the screw was tightened, was deformed as a result of the tensile test. In the DIC tensile test, Part 1 showed the localized deformation until fracture occurred, and Part 3 was slightly deformed compared to Part 1, whereas Part 2 sustained against to deformation. Similarly, as a result of FEM analysis with two lines (**Fig. 8B**), the equivalent strain of the sensor embedded region was less than 0.01 (**Fig. 9**).

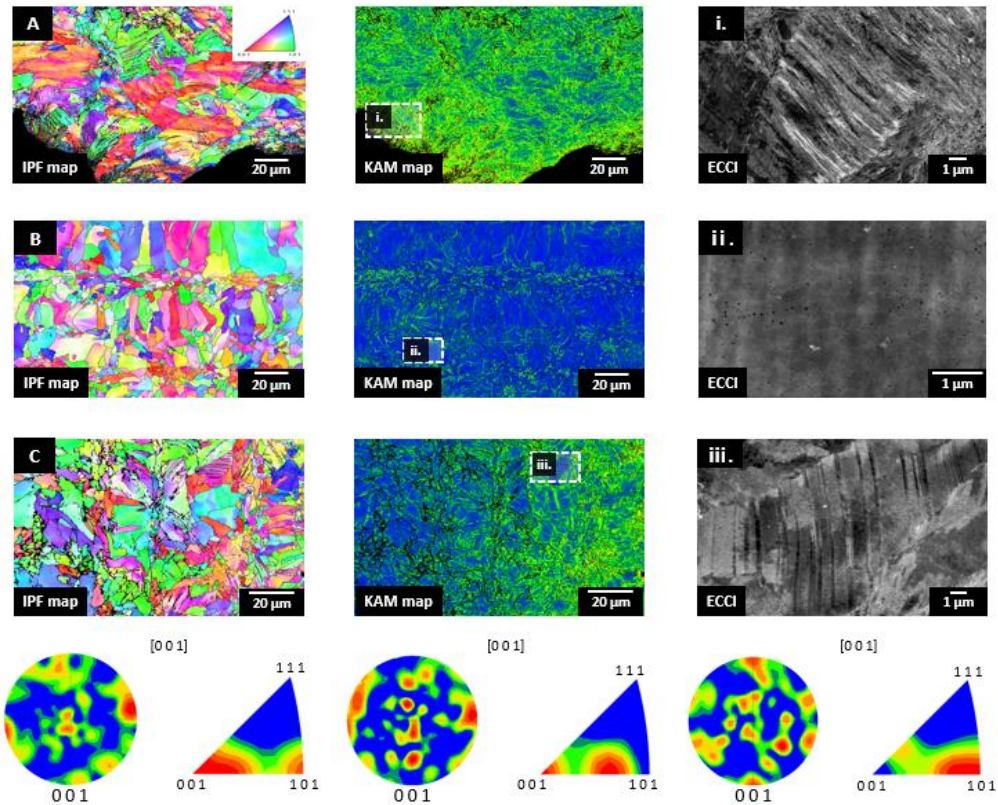

**Fig. 10. Inverse pole figure (IPF), Kernel average misorientation (KAM), and electron channeling contrast imaging (ECCI) analysis in each location of tensile tested specimen** (A) in the fractured region, (B) sensor embedded area, and (C) vicinity of hole.

In addition to these analyses, microstructural analysis was also conducted to confirm the stability of sensor embedded part (**Fig. 6**). EBSD, inverse pole figure (IPF), kernel average misorientation (KAM), and ECCI analysis were performed to analyze the material properties of each major region. The IPF and texture analysis revealed that {001} and {101} planes were more dominant rather than the {111} plane owing to the epitaxial heat release through the <001> and <101> directions. According to the KAM analysis, the local grain misorientation value was high in the highly deformed area. According to the KAM analysis, the local grain misorientation value was high in the highly deformed area. As shown in **Fig. 6A**, tensile failure occurred near the hole, and a typical ductile fracture mode was observed. Most of this area was heavily deformed with high KAM values, and a large amount of twin deformation was also observed with a misorientation angle of 60°. Interestingly, the sensor-embedded area (**Fig. 6B**) was little deformed despite of the empty internal space; this indicates that safety can be ensured in the sensor-embedded area. As shown in **Fig. 6C**, densely deformed were located in the vicinity of the hole owing to a high stress concentration factor (Kt).

*3.2 Analysis of loose screws*

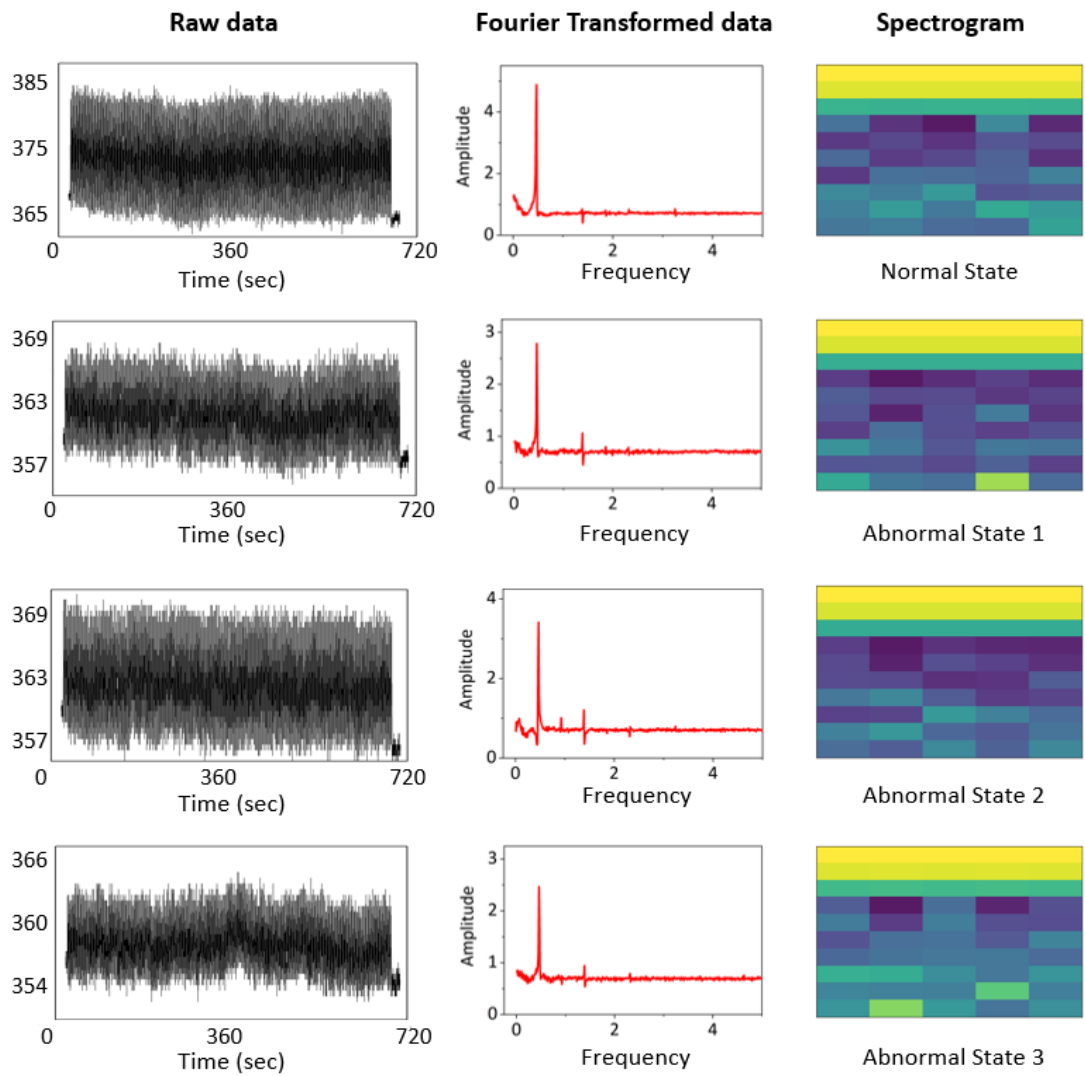

**Fig. 11. Data preprocessing of state classification test.** The deformation data of the repeated compression test were preprocessed through FFT to spectrograms.

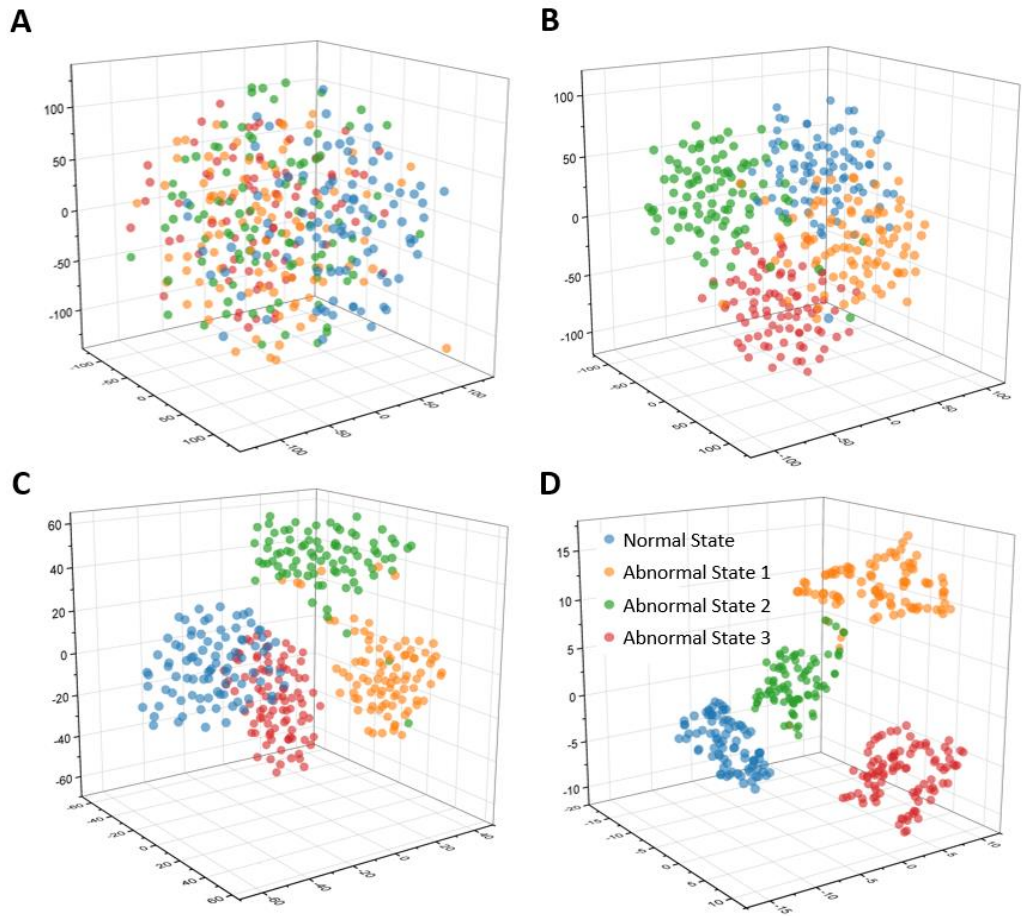

**Fig. 12. AI-augmented metal malfunction state classification process (normal state – abnormal states 1, 2, 3).** Optimization process of CNN t-SNE classification for metal component status classification test: (A) GradientDescent optimizer & learning rate: 0.0002, (B) Adagrad optimizer & learning rate: 0.002, (C) Adam optimizer & learning rate: 0.04, (D) Adam optimizer & learning rate: 0.02

The repetitive compression tests were conducted by dividing the normal state and three abnormal states. **Fig. 11** shows the metal component's malfunction state classification test and the process of converting data from repeated compression tests into spectrograms. The strain gauge embedded inside the metal component printed strain data. The strain data output for 720 seconds was used after removing the inaccurate data of the last 20 seconds. As shown in the raw data plots, the 2nd and 4th experiments showed different data characteristics for each of the 4 tests, such as a decrease in amplitude, but they did not show a clear difference that could distinguish the states. First, we used FFT to convert raw data into a frequency domain. As the data could not be differentiated easily, it was difficult to create a spectrogram with only raw data and learn it using the CNN; thus, FFT-based data were used. The FFT algorithm can quickly calculate the discrete Fourier transform (DFT) that decomposes discrete input signals into frequencies. The representative definition of DFT is as follows.

$$f_j = \sum_{k=0}^{n-1} x_k e^{-\frac{2\pi i}{n} jk}, \qquad j = 0, \ldots, n-1 \qquad (1)$$

The time-domain data was output as a complex number consisting of real and imaginary values (amplitude and phase) and was moved to the frequency domain. As a result, the time–strain data of the metal part were converted to frequency–amplitude data via the FFT every 6 s. However, although there were differences among the data, it was still not appropriate to learn with the CNNs because of the blank space in the graph images to learn, so we used a strategy to convert data into spectrograms. The data converted to the frequency domain were converted again to a spectrogram for the CNN. Spectrograms were created every 6 s with an image size of 10×5 and were used as input data for the CNN. Our CNN learned the labeled spectrograms, and the results were visualized in 3D space. **Fig. 12** shows the CNN model optimization process of t-SNE classification with changes in learning rate and optimizer. We adjusted the learning rate and type of optimizer to create the neural network with the lowest training/validation cost without overfitting the data as shown in **Fig. S8**. The results indicate that the validation cost did not overtake the training cost and neither over or under fitting occurred in the CNN model. As shown in **Table 1,** the optimized CNN model has a learning rate of 0.02 and the optimizer is 'Adam'. Raw data obtained from the repeated compression tests were generated as 424 spectrograms, of which 85%, or 360, were used as the training sets, and the remaining 15%, or

64, were used as the test sets. Our network showed a 98.6% accuracy (355 out of 360) on the training sets. An accuracy of 89.06% (57 out of 64) was obtained on the test sets. In the prediction results, of both the training sets and the test sets, there were 5 and 7 false predictions, respectively. All false predictions were cases of classification between abnormal state 1 and abnormal state 2. However, it is noteworthy that there was no false prediction in distinguish between normal and abnormal states.

**Table 1. Summary of CNN optimization process with optimizer & learning rate (Status classification test)**

|  | **Case 1** | **Case 2** | **Case 3** | **Opt. Case** |
| --- | --- | --- | --- | --- |
| **Optimizer** | Gradient Descent | Adagrad | Adam | Adam |
| **Learning rate** | 0.0002 | 0.002 | 0.04 | 0.02 |
| **Cost** | 1.3794 | 1.3774 | 0.9407 | 0.7841 |

*3.3 Analysis of the cause of dynamic vibrations*

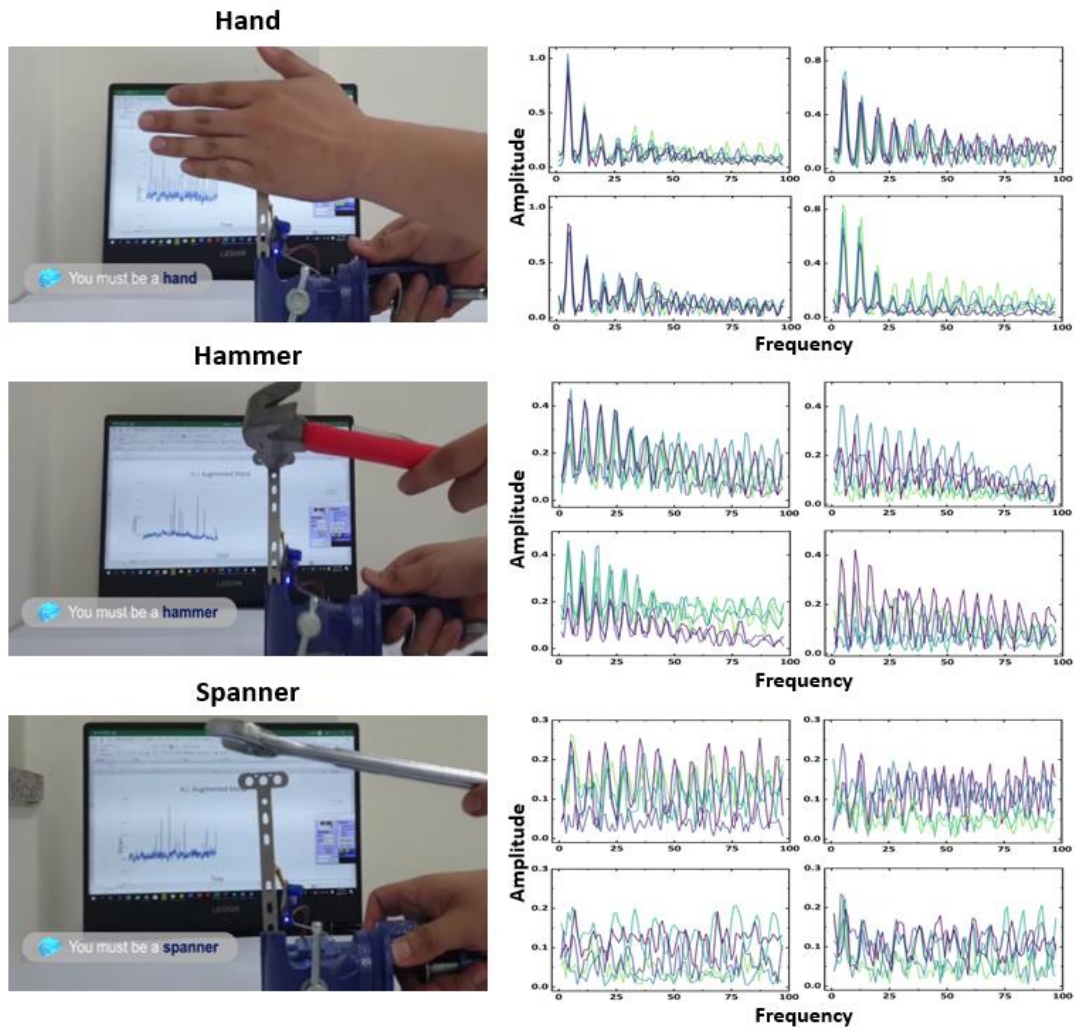

**Fig. 13. Real-time recognition test of external impact objects and data preprocessing with FFT.** AI-augmented metal distinguishes three externally impacted objects in real time. The real time deformation data were converted through FFT.

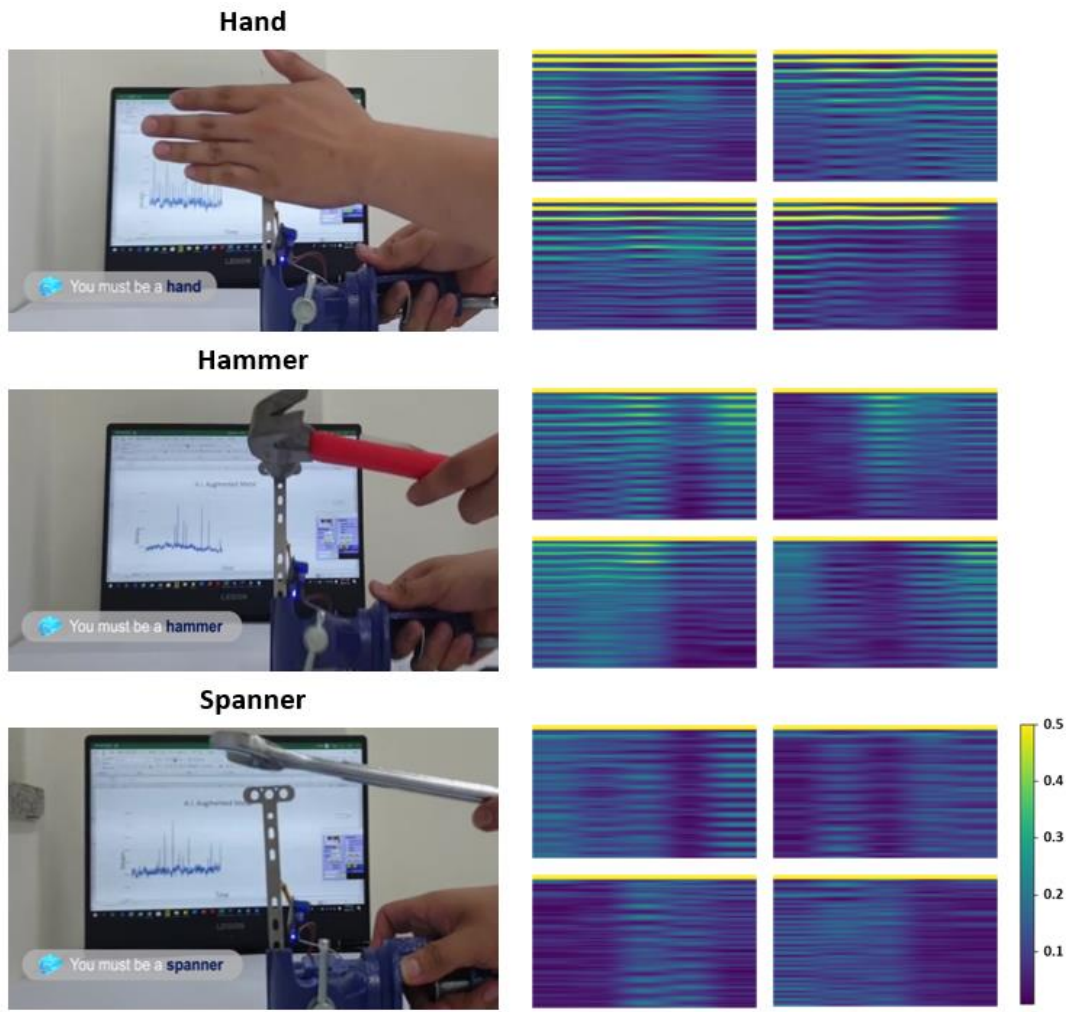

**Fig. 14. Real-time recognition test of external impact objects and converting process of deformation data into spectrogram.** Each FFT data was converted into a spectrum and then labeled according to type of impact objects.

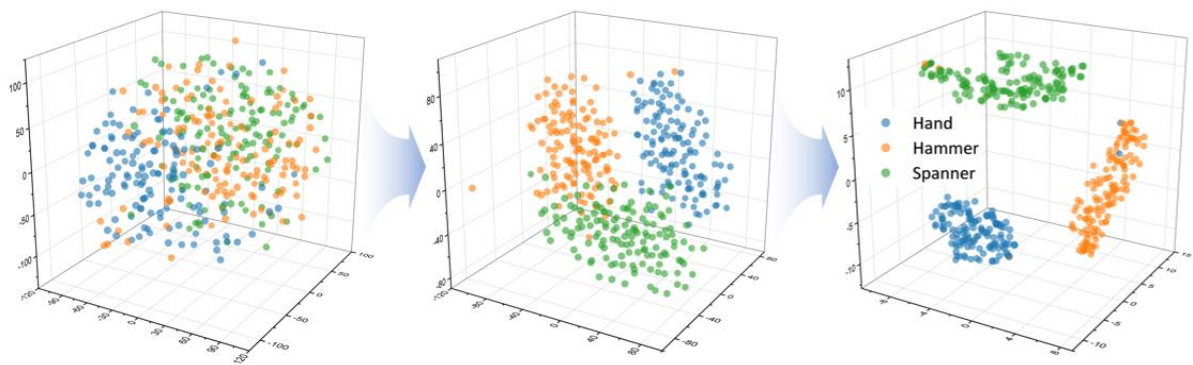

**Fig. 15. Optimization process of t-SNE classification for real-time recognition test.** GradientDescent optimizer & learning rate: 0.0002, Adagrad optimizer & learning rate: 0.002, Adam optimizer & learning rate: 0.02.

The AIADM component was also demonstrated to verify unknown external object impact (**Fig. 13**). The external objects used in the experiment were a hand, a hammer, and a spanner. The metal learned 10 minutes of strain data generated by the impact of external objects. After the learning, we tested the AI model to distinguish between external objects by tapping it in real time. The result of the AI classification of the metal component was output as sound through the speaker (Supplementary Video 1, 2). The strain data generated by the metal component was Fourier fast transformed and then converted into a spectrogram for CNN training (**Fig. 14**). To visualize the results of recognition test, a t-SNE 3D scatter plot was used again. **Fig. 15** and **Table 2** show the CNN model optimization process of AI recognition test with changes in learning rate and optimizer. The optimized CNN model (Adam optimizer, learning rate 0.02, cost value 0.55) showed an accuracy of 97.34% (402 out of 413) on the training set (85% of the data) and an accuracy of 83.56% (61 out of 73) on the test set (15% of the data).

**Table 2. Summary of CNN optimization process with optimizer & learning rate (Real-time recognition test of external impact objects)**

|  | Case 1 | Case 2 | Opt. Case |
|---|---|---|---|
| **Optimizer** | Gradient Descent | Adagrad | Adam |
| **Learning rate** | 0.0002 | 0.002 | 0.02 |
| **Cost** | 1.0915 | 1.0741 | 0.5517 |

**IV. CONCLUSIONS**

Here, we introduce a brilliant metal which can accurately verify the subtle external change. Using L-PBF, we embedded a strain gauge in the desired part of the mechanical system without deformation of metal components. Strain gauge was successfully embedded inside metal based on L-PBF, and CNN-based AI technique was used to determine the state of the metal part based on the sensor data. The results indicate that the local release of screws or the positions of loose screws can be determined. In addition, the type of unknown external impacting object can be distinguished. In this study, we demonstrated simple t-shape AI augmented digital, but this work can be extended with diverse 3D shape metal components for various applications which our developed fabrication process and optimized CNN process. In addition, while we employed wired metal system, large scale metal component with various wireless communication modules such as Wi-Fi or Bluetooth can be embedded with our developed process for ultimate hyperconnectivity of metal system. We expect that AIADM will pave the way for next generation of digital metal based mechanical systems, smart factories, autonomous vehicles, or robots.

**CRediT authorship contribution statement**

**Eunhyeok Seo, Hyokyung Sung:** Formal analysis, Visualization, Writing – original draft, Writing – review & editing, Investigation**,** Conceptualization. **Taekyeong Kim, Sangeun Park, Min Sik Lee, Jung Gi Kim:** Formal analysis, Investigation. **Ji-hun Yu, Kyung Tae Kim:** Visualization, Investigation. **Hayoung Chung, Seong Jin Park, Namhun Kim:** Methodology. **Hayeol Kim, Seung Ki Moon, Seong-Kyum Choi:** Visualization, Writing – review & editing. **Im Doo Jung:** Conceptualization, Formal analysis, Supervision, Funding acquisition, Writing – original draft, Writing – review & editing

## DECLARATION OF COMPETING INTEREST

The authors declare that they have no known competing financial interests or personal relationships that could have appeared to influence the work reported in this paper.

## ACKNOWLEDGMENTS

This work was supported by the National Research Foundation of Korea (NRF) grant funded by the Korea government (MSIT) (grant nos. 2021M2D2A1A01050059 and 2021R1F1A1046079).

# Supplementary material

**AI Augmented Digital Metal Component**

Eunhyeok Seo[a]*, Hyokyung Sung[b]*, Hayeol Kim[a], Taekyeong Kim[a], Sangeun Park[b],

Min Sik Lee[a], Seung Ki Moon[c], Jung Gi Kim[b], Hayoung Chung[a], Seong-Kyum Choi[d], Ji-hun

Yu[e], Kyung Tae Kim[e], Seong Jin Park[f], Namhun Kim[a], Im Doo Jung[a]†

[a]Department of Mechanical Engineering, Ulsan National Institute of Science and Technology, Ulsan 44919, Republic of Korea

[b]Department of Materials Engineering and Convergence Technology, Gyeongsang National University, Jinju 52828, Republic of Korea

[c]School of Mechanical and Aerospace Engineering, Nanyang Technological University, 50 Nanyang Ave, 639798, Singapore

[d]G.W.W. School of Mechanical Engineering, Georgia Institute of Technology, Atlanta, GA 30332, USA

[e]Powder/Ceramic Materials Division, Korea Institute of Materials Science, Changwon 51508, Republic of Korea

[f]Department of Mechanical Engineering, Pohang University of Science and Technology, Pohang 37673, Republic of Korea

†Corresponding author: Prof. Im Doo Jung (idjung@unist.ac.kr)

*These authors contributed equally to this work.

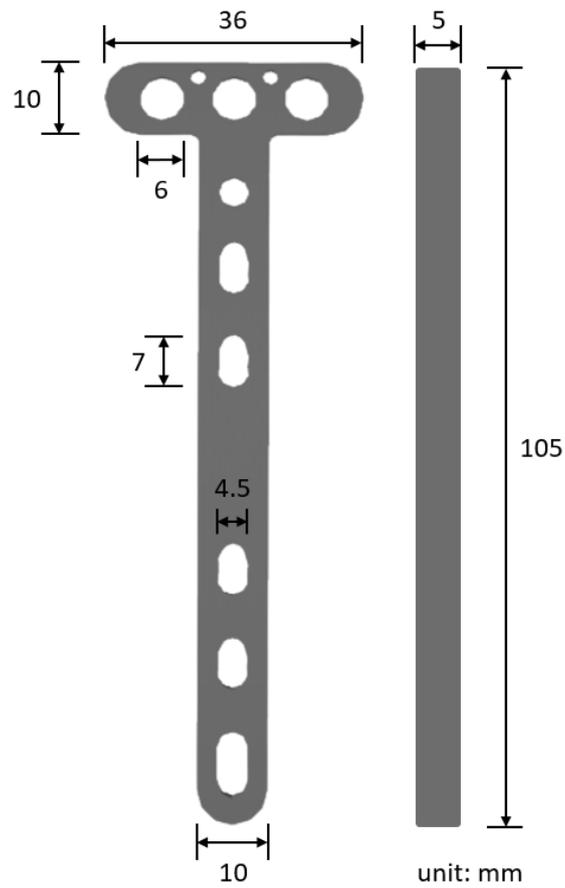

**Fig. S1. Dimension of T-shaped sensor embedded part geometry.** The dimensions of T-shaped part geometry are 105mm x 36mm x 5mm; length of 105mm, width of 35mm (wide) and 10mm (narrow), and thickness of 5 mm. Three holes in head part have 6mm, and the length of the major axis of the elliptical hole is 7 mm and the minor axis is 4.5 mm.

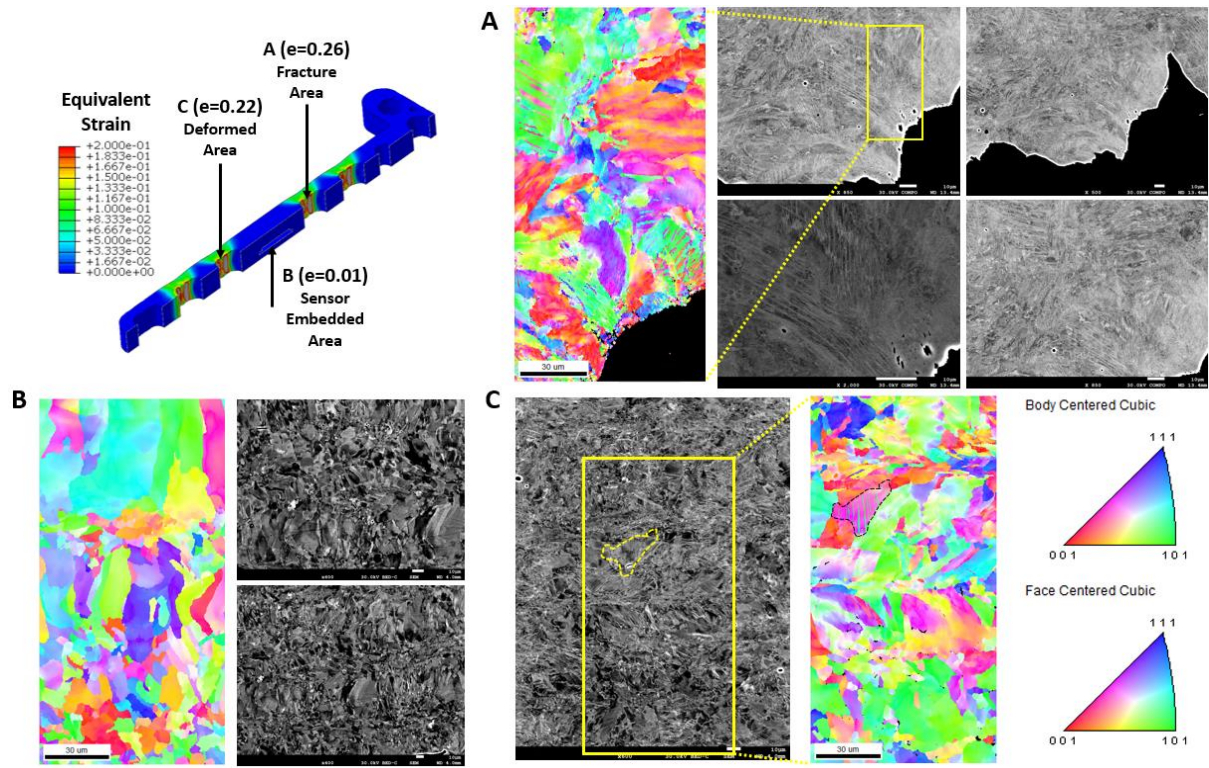

**Fig. S2. FEM analysis results, EBSD analysis images, and ECCI low scale analysis results.** (A) fracture area, (B) sensor embedded area, and (C) deformation area. In the result of FEM analysis, the equivalent strain was 0.26 and 0.22 for parts A and C, respectively, but the strain for part B was 0.01.

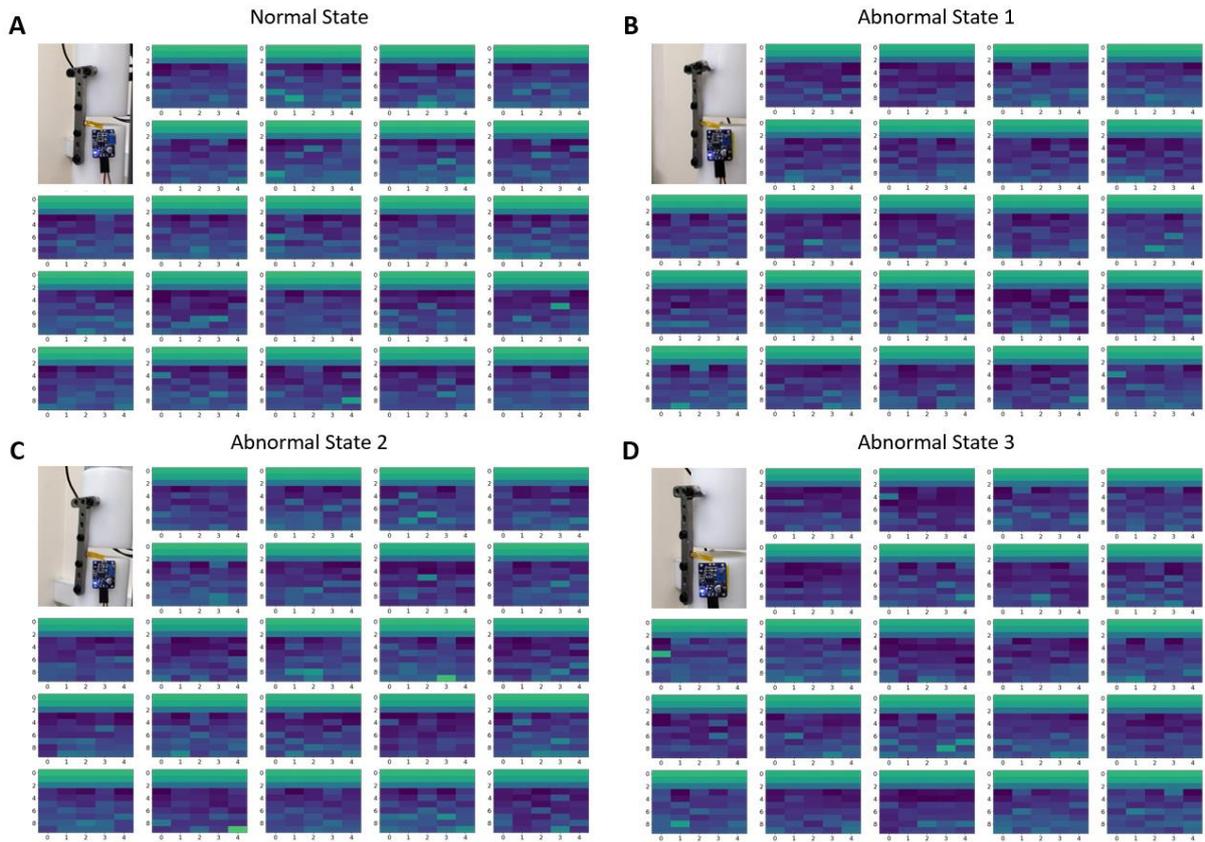

**Fig. S3. Spectrograms of 4 different AIADM component statements in repeated compression test.** The strain data for 4 cases, (A) normal state, (B) abnormal state 1 (loose left screw), (C) abnormal state 2 (left screw missing), and (D) abnormal state 3 (left screw missing & loose right screw), set in the repeated compression test were converted into 106 spectrograms, respectively. These images were used as training data in CNN learning.

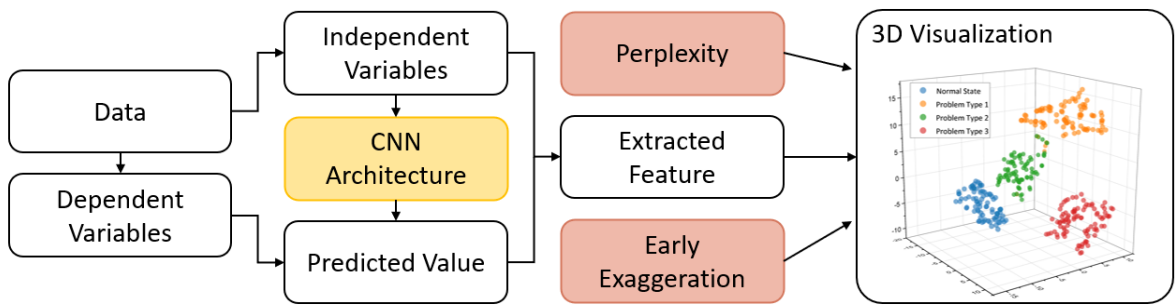

**Fig. S4. Flow chart of t-SNE visualization.** We used Scikit learn's t-SNE tool for t-SNE visualization. As parameters, perplexity value 13 and early exaggeration value 4 were applied to classify the states of AIADM component in 3D space.

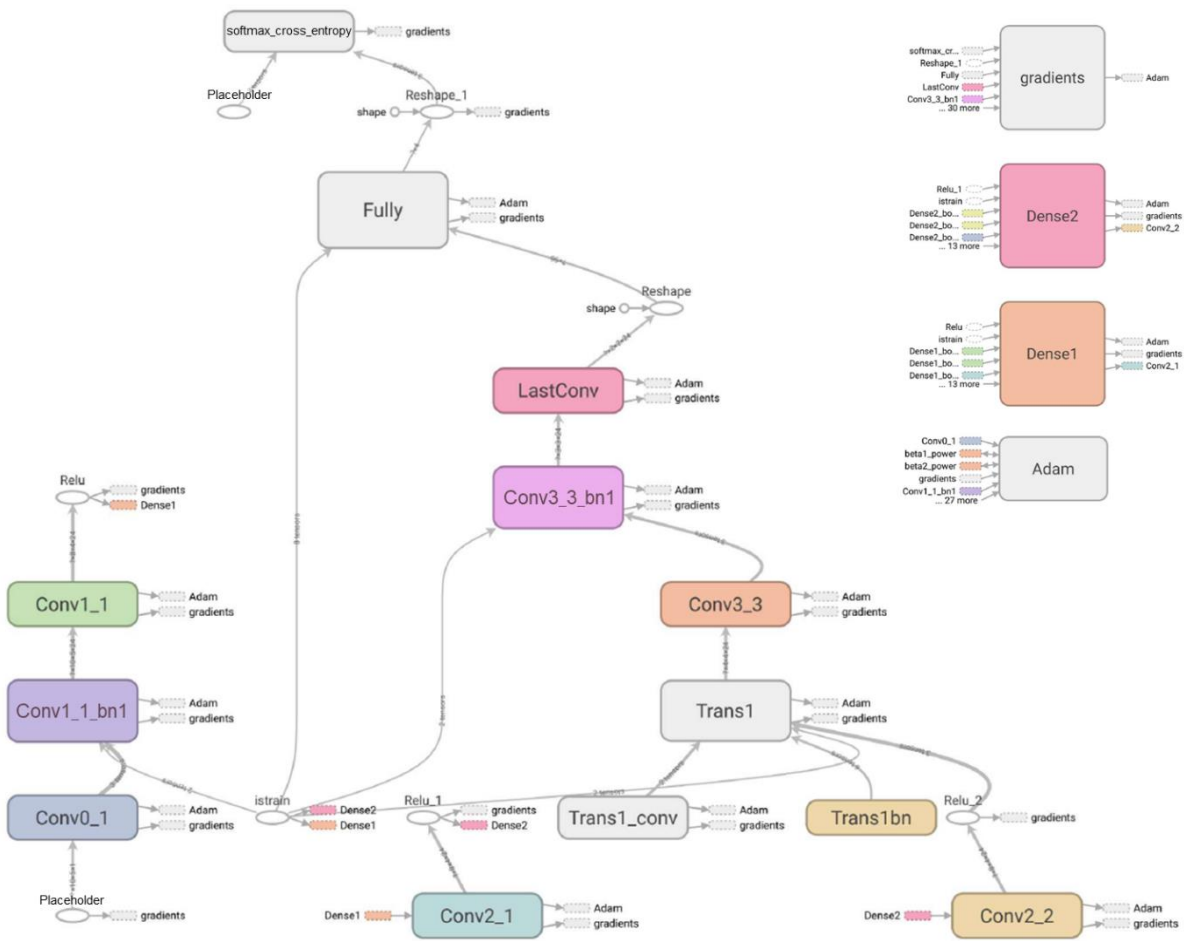

**Fig. S5. Flow chart of CNN presented with Jupyter notebook.** The CNN structure was visualized using Tensorboard. This CNN structure is consists of multiple convolutional layers and dense layers.

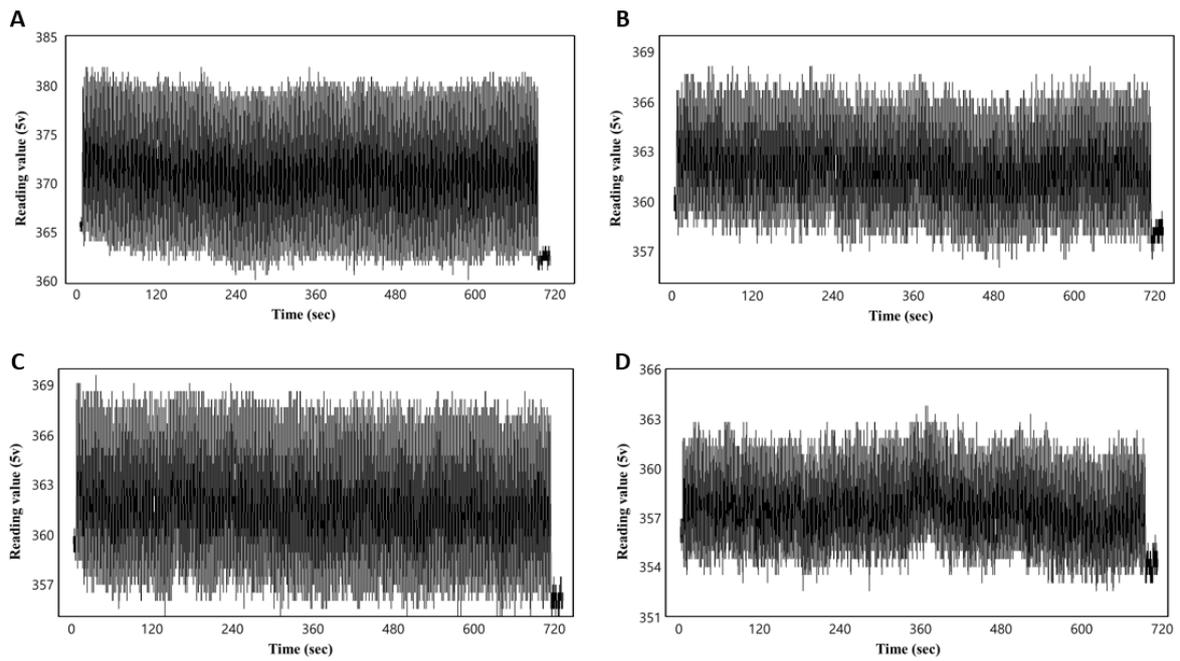

**Fig. S6. Time series graph of AIADM deformation data (state classification test).** (A) normal state (B) abnormal state1: left screw is loose (C) abnormal state 2: left screw is missing (D) abnormal state 3: the left screw is loose, and the right screw is missing.

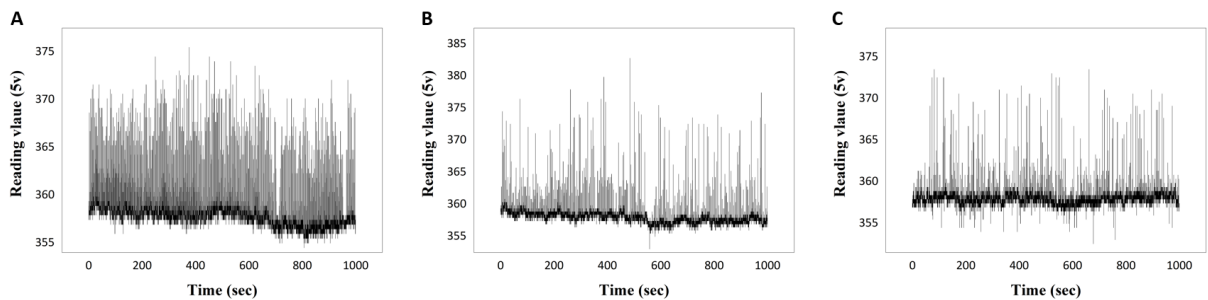

**Fig. S7. Time series graph of AIADM deformation data (impact recognition test).** (A) Hand (B) Hammer (C) Spanner

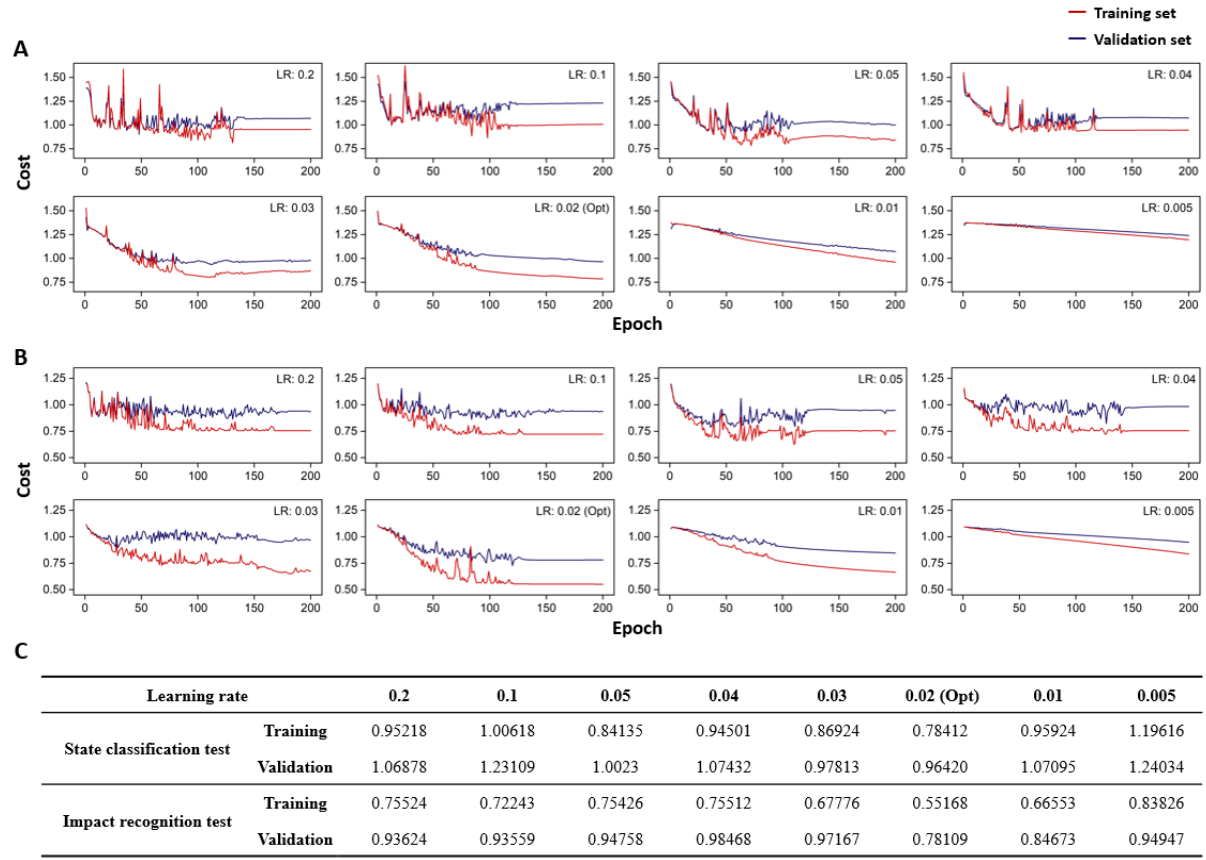

**Fig. S8. Cost graphs with diverse learning rates and change of epoch.** In both cases, (A) state classification test and (B) impact recognition test showed (C) lowest cost value when the learning rate was 0.02.

**Table S1. Deformation data of repeated compression test (AIADM state classification test)**

| Time (sec) | Normal State | Abnormal State 1 | Abnormal State 2 | Abnormal State 3 |
|---|---|---|---|---|
| 0.1 | 365.72 | 359.86 | 359.86 | 355.47 |
| 0.2 | 366.21 | 359.38 | 359.38 | 355.47 |
| 0.3 | 365.72 | 360.35 | 359.38 | 355.96 |
| 0.4 | 366.21 | 360.35 | 359.86 | 356.45 |
| 0.5 | 366.21 | 359.38 | 359.86 | 356.45 |
| 0.6 | 366.21 | 360.84 | 359.38 | 355.96 |
| 0.7 | 365.23 | 359.86 | 358.89 | 356.45 |
| 0.8 | 366.21 | 360.35 | 359.86 | 356.45 |
| 0.9 | 366.21 | 360.84 | 359.86 | 356.93 |
| 1 | 366.21 | 359.86 | 359.38 | 355.96 |
| 1.1 | 365.72 | 360.35 | 359.38 | 356.45 |
| 1.2 | 365.72 | 360.35 | 360.35 | 356.45 |
| 1.3 | 365.72 | 360.35 | 359.38 | 356.45 |
| 1.4 | 366.21 | 359.86 | 359.38 | 355.96 |
| 1.5 | 366.21 | 360.35 | 359.86 | 356.93 |
| 1.6 | 365.72 | 360.35 | 359.38 | 355.96 |
| 1.7 | 365.72 | 359.38 | 358.89 | 355.47 |
| 1.8 | 366.7 | 359.38 | 359.38 | 355.96 |
| ... | ... | ... | ... | ... |
| 664.1 | 363.28 | 358.4 | 356.45 | 354.49 |
| 664.2 | 363.28 | 358.4 | 355.96 | 354 |
| 664.3 | 362.79 | 358.4 | 355.47 | 354 |
| 664.4 | 362.79 | 357.91 | 355.47 | 354.49 |
| 664.5 | 363.28 | 358.89 | 355.47 | 354.98 |
| 664.6 | 362.79 | 356.93 | 355.47 | 354.98 |
| 664.7 | 362.79 | 357.91 | 355.47 | 353.52 |
| 664.8 | 362.79 | 357.91 | 355.47 | 354 |
| 664.9 | 363.28 | 358.4 | 355.47 | 354.98 |
| 665 | 362.79 | 358.4 | 355.47 | 354 |
| 665.1 | 362.79 | 358.4 | 355.47 | 354 |
| 665.2 | 362.3 | 358.4 | 355.47 | 353.52 |
| 665.3 | 362.3 | 357.91 | 355.96 | 354.98 |
| 665.4 | 362.3 | 358.4 | 355.47 | 354 |
| 665.5 | 362.3 | 358.4 | 356.45 | 354.49 |
| 665.6 | 362.3 | 358.4 | 355.96 | 354 |
| 665.7 | 362.79 | 358.4 | 355.96 | 354 |
| 665.8 | 362.3 | 358.4 | 356.45 | 354.49 |
| 665.9 | 361.82 | 358.4 | 355.96 | 354 |

**Table S2. Deformation data of AIADM impact recognition test**

| Time (sec) | Hand | Hammer | Spanner |
|---|---|---|---|
| 0.1 | 359.38 | 358.40 | 356.93 |
| 0.2 | 359.38 | 358.40 | 357.42 |
| 0.3 | 358.89 | 358.40 | 357.42 |
| 0.4 | 359.38 | 358.89 | 356.93 |
| 0.5 | 359.38 | 358.89 | 357.91 |
| 0.6 | 358.40 | 359.38 | 357.42 |
| 0.7 | 357.42 | 359.38 | 357.42 |
| 0.8 | 357.91 | 358.89 | 357.42 |
| 0.9 | 358.89 | 358.40 | 358.40 |
| 1 | 358.40 | 358.89 | 358.40 |
| 1.1 | 358.89 | 358.40 | 357.91 |
| 1.2 | 357.91 | 358.89 | 357.91 |
| 1.3 | 357.91 | 359.38 | 357.91 |
| 1.4 | 358.40 | 359.38 | 356.93 |
| 1.5 | 357.91 | 358.89 | 357.42 |
| 1.6 | 357.91 | 358.89 | 357.91 |
| 1.7 | 358.40 | 358.89 | 357.42 |
| 1.8 | 357.91 | 359.38 | 356.93 |
| ... | ... | ... | ... |
| 998.2 | 357.42 | 357.91 | 357.42 |
| 998.3 | 356.93 | 357.42 | 356.93 |
| 998.4 | 357.42 | 357.91 | 356.93 |
| 998.5 | 356.45 | 357.91 | 356.93 |
| 998.6 | 357.42 | 357.91 | 356.45 |
| 998.7 | 356.93 | 358.89 | 356.45 |
| 998.8 | 356.45 | 356.93 | 357.42 |
| 998.9 | 356.45 | 357.42 | 357.42 |
| 999 | 355.96 | 357.91 | 357.42 |
| 999.1 | 356.93 | 357.91 | 357.42 |
| 999.2 | 355.96 | 357.42 | 356.93 |
| 999.3 | 356.45 | 357.91 | 357.42 |
| 999.4 | 356.93 | 357.91 | 356.93 |
| 999.5 | 357.42 | 356.93 | 356.93 |
| 999.6 | 356.93 | 356.45 | 356.93 |
| 999.7 | 357.42 | 356.93 | 356.93 |
| 999.8 | 357.91 | 356.93 | 357.42 |
| 999.9 | 357.42 | 356.93 | 357.42 |

**Table S3. Training set of AIADM state classification test**

| Raw data | Spectrogram | Ground Truth | Prediction |
|---|---|---|---|
| 363.77<br>362.30<br>…<br>363.77<br>365.72 | 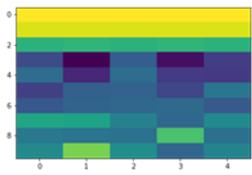 | Abnormal State2 | Abnormal State2 |
| 375.49<br>376.95<br>…<br>365.23<br>367.19 | 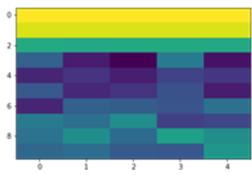 | Normal State | Normal State |
| 368.16<br>370.61<br>…<br>366.70<br>364.75 | 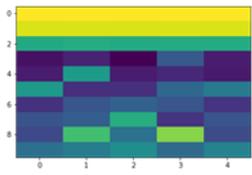 | Normal State | Normal State |
| 358.89<br>359.86<br>…<br>361.82<br>361.33 | 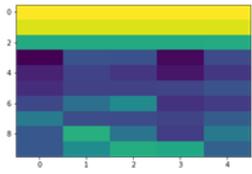 | Abnormal State1 | Abnormal State1 |
| … | … | … | … |
| 375.49<br>378.42<br>…<br>365.23<br>367.68 | 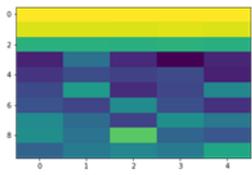 | Normal State | Normal State |
| 360.35<br>359.86<br>…<br>367.19<br>366.21 | 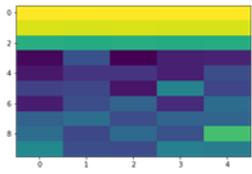 | Abnormal State1 | Abnormal State1 |
| 357.91<br>356.93<br>…<br>363.28<br>362.30 | 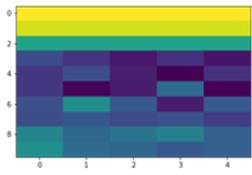 | Abnormal State2 | Abnormal State2 |
| 360.84<br>361.33<br>…<br>360.35<br>358.89 | 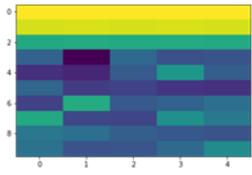 | Abnormal State3 | Abnormal State3 |

**Table S4. Validation set of AIADM state classification test**

| Raw data | Spectrogram | Ground Truth | Prediction |
|---|---|---|---|
| 357.42<br>357.91<br>…<br>360.84<br>361.33 | 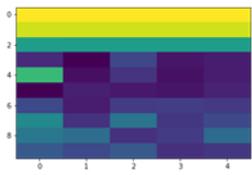 | Abnormal State3 | Abnormal State3 |
| 359.86<br>359.38<br>…<br>365.23<br>363.77 | 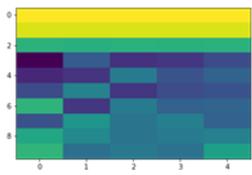 | Abnormal State1 | Abnormal State1 |
| 363.28<br>362.30<br>…<br>364.75<br>365.72 | 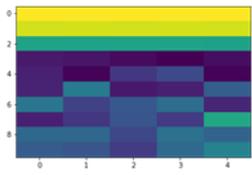 | Abnormal State1 | Abnormal State1 |
| 361.33<br>360.35<br>…<br>366.70<br>368.16 | 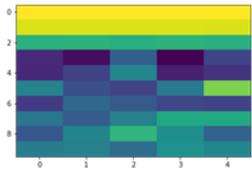 | Abnormal State2 | Abnormal State2 |
| … | … | … | … |
| 359.38<br>357.91<br>…<br>366.21<br>364.75 | 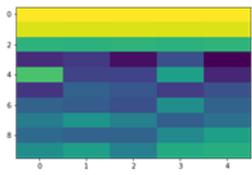 | Abnormal State2 | Abnormal State2 |
| 354.98<br>354.98<br>…<br>355.47<br>354.49 | 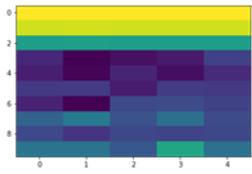 | Abnormal State3 | Abnormal State3 |
| 370.12<br>369.14<br>…<br>380.37<br>378.91 | 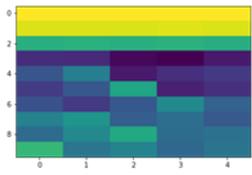 | Normal State | Normal State |
| 377.44<br>375.00<br>…<br>371.58<br>373.54 | 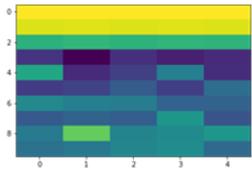 | Normal State | Normal State |

**Table S5. Test set of AIADM state classification test**

| Raw data | Spectrogram | Ground Truth | Prediction |
|---|---|---|---|
| 359.38<br>359.86<br>…<br>364.26<br>362.30 | 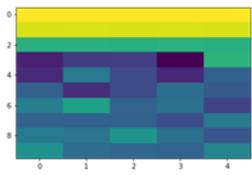 | Abnormal State1 | Abnormal State1 |
| 366.70<br>364.75<br>…<br>363.77<br>365.23 | 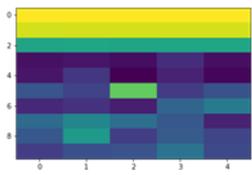 | Abnormal State2 | Abnormal State1 |
| 367.19<br>365.72<br>…<br>377.44<br>375.98 | 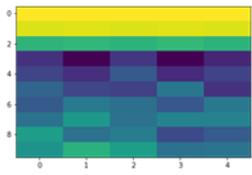 | Normal State | Normal State |
| 355.96<br>354.49<br>…<br>359.86<br>358.89 | 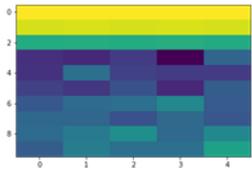 | Abnormal State3 | Abnormal State3 |
| … | … | … | … |
| 356.45<br>356.45<br>…<br>355.47<br>354.49 | 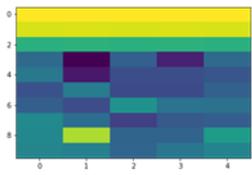 | Abnormal State3 | Abnormal State3 |
| 354.98<br>354.00<br>…<br>358.89<br>358.40 | 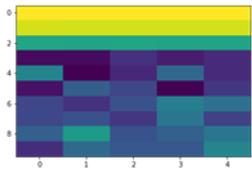 | Abnormal State3 | Abnormal State3 |
| 379.39<br>376.46<br>…<br>370.61<br>372.56 | 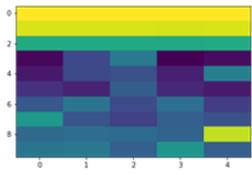 | Normal State | Normal State |
| 359.38<br>359.86<br>…<br>360.35<br>359.86 | 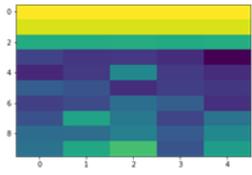 | Abnormal State1 | Abnormal State2 |

**Table S6. Training set of AIADM impact recognition test**

| Raw data | Spectrogram | Ground Truth | Prediction |
|---|---|---|---|
| 359.38<br>357.91<br>…<br>357.91<br>358.40 | 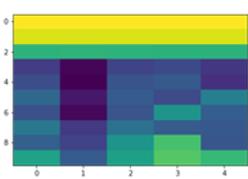 | Spanner | Spanner |
| 358.40<br>357.91<br>…<br>360.84<br>357.91 | 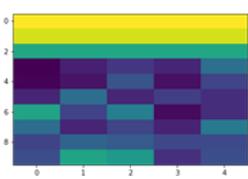 | Hammer | Hammer |
| 358.40<br>358.40<br>…<br>358.40<br>358.40 | 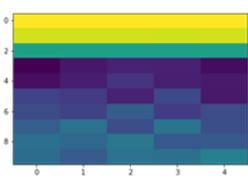 | Hammer | Hammer |
| 357.42<br>357.91<br>…<br>366.21<br>361.82 | 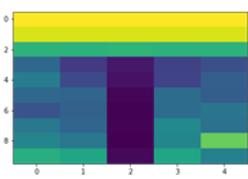 | Hand | Hand |
| … | … | … | … |
| 356.45<br>357.42<br>…<br>357.42<br>357.91 | 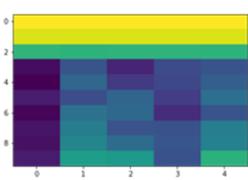 | Hand | Hand |
| 356.93<br>356.45<br>…<br>356.45<br>356.45 | 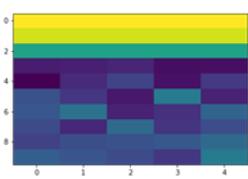 | Spanner | Spanner |
| 357.42<br>357.91<br>…<br>357.42<br>357.42 | 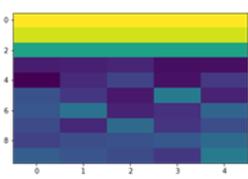 | Hand | Hand |
| 357.42<br>356.93<br>…<br>357.91<br>358.40 | 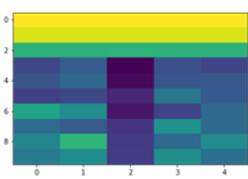 | Hammer | Hammer |

**Table S7. Validation set of AIADM impact recognition test**

| Raw data | Spectrogram | Ground Truth | Prediction |
|---|---|---|---|
| 358.40<br>358.40<br>…<br>358.40<br>367.19 | 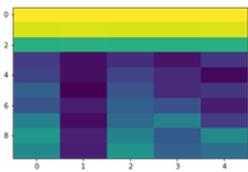 | Hand | Hand |
| 358.40<br>358.89<br>…<br>358.40<br>358.89 | 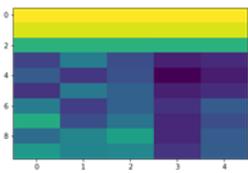 | Hammer | Hammer |
| 365.23<br>362.30<br>…<br>360.35<br>360.35 | 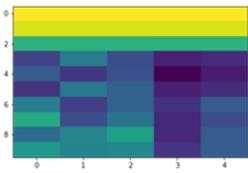 | Hand | Hand |
| 357.91<br>357.91<br>…<br>357.42<br>358.40 | 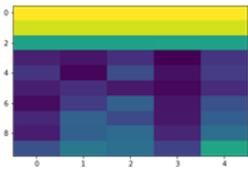 | Spanner | Spanner |
| … | … | … | … |
| 358.89<br>358.40<br>…<br>357.91<br>357.91 | 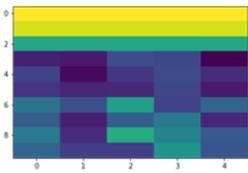 | Hand | Hand |
| 357.91<br>358.40<br>…<br>358.89<br>359.38 | 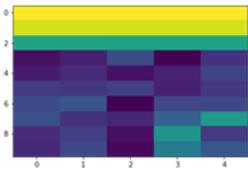 | Hammer | Hammer |
| 358.40<br>358.89<br>…<br>358.40<br>358.40 | 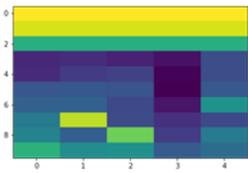 | Hammer | Hammer |
| 358.89<br>358.40<br>…<br>358.40<br>358.40 | 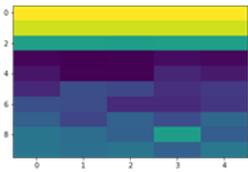 | Hand | Hand |

**Table S8. Test set of AIADM impact recognition test**

| Raw data | Spectrogram | Ground Truth | Prediction |
|---|---|---|---|
| 357.91<br>358.40<br>…<br>358.89<br>358.89 | | Hammer | Hammer |
| 357.91<br>358.89<br>…<br>359.86<br>359.86 | | Hammer | Hammer |
| 359.38<br>366.70<br>…<br>368.16<br>362.79 | | Hand | Hammer |
| 360.84<br>361.33<br>…<br>363.77<br>363.28 | | Hand | Hand |
| … | … | … | … |
| 358.89<br>358.89<br>…<br>358.40<br>357.91 | | Spanner | Spanner |
| 357.91<br>357.42<br>…<br>356.93<br>356.93 | | Spanner | Spanner |
| 356.93<br>357.42<br>…<br>358.40<br>358.40 | | Spanner | Hammer |
| 357.91<br>357.91<br>…<br>356.93<br>357.91 | | Spanner | Hammer |

**Full data can be found in the attached file below.**

[Dataset of AIADM state classification test.xlsx](Dataset of AIADM state classification test.xlsx)

[Dataset of AIADM impact recognition test.xlsx](Dataset of AIADM impact recognition test.xlsx)